\documentclass[aps,pra,twocolumn,groupedaddress,10pt]{revtex4}

\usepackage{amsmath}
\usepackage{amsfonts}
\usepackage{amssymb}
\usepackage{graphicx}
\usepackage{hyperref}

\hypersetup{colorlinks=true, urlcolor=blue, linkcolor=blue, citecolor=blue,plainpages=false}

\newcommand{\txtpow}[1]{{\mbox{\scriptsize{#1}}}}

\begin{document}
\title{Level repulsion in hybrid photonic-plasmonic microresonators for enhanced biodetection}

\author{Matthew R. Foreman}
\email[]{matthew.foreman@mpl.mpg.de}
\author{Frank Vollmer}
\affiliation{Max Planck Institute for the Science of Light,
Laboratory of Nanophotonics and Biosensing, 
G\"unther-Scharowsky-Stra{\ss}e 1, 91058 Erlangen, Germany}

\begin{abstract}
We theoretically analyse photonic-plasmonic coupling
between a high $Q$ whispering gallery mode (WGM) resonator and a
core-shell nanoparticle.  Blue and red shifts of WGM resonances are
shown to arise from crossing of the photonic and plasmonic
modes. Level repulsion in the hybrid system is further seen to
enable sensitivity enhancements in WGM sensors: maximal when the two
resonators are detuned by half the plasmon linewidth. Approximate bounds are given to quantify
possible enhancements. Criteria for reactive vs.
resistive coupling are also established. \\ 
\\
DOI:
\href{dx.doi.org/10.1103/PhysRevA.88.023831}{10.1103/PhysRevA.88.023831}
\hfill PACS number(s): 42.79.Gn, 73.20.Mf, 78.67.Bf, 87.80.Nj
\end{abstract}

%


\maketitle

\section{Introduction}

Optical microresonators play an important role in modern day physics,
for example, enabling study of cavity quantum electrodynamic 
phenomena, tailoring of spontaneous emission spectra of atoms and
quantum dots, spectral filtering and development of novel light
sources \cite{Vahala2003}.
Combination of optical microresonators with metallic nanoparticles
(NPs) supporting localised surface plasmons (LSPs)
has also recently attracted much attention, due to the opportunities
it affords in efficient light routing, field confinement and
enhanced spectroscopy \cite{Ahn2012,Chamanzar2011,White2007,Xiao2012}. One further
field of importance is that of whispering gallery mode (WGM)
biosensing where large reactive coupling to dielectric particles, such
as bacteria and proteins, results in detectable WGM frequency shifts \cite{Ren2007, Noto2007}. In the
drive for single molecule sensitivity, enhancement of
reactive shifts and maximisation of near field intensities are, however, necessary
\cite{Vollmer2008} such that plasmonic NPs
are being increasingly employed as either analyte labels \cite{Lin2013,Santiago2011,Witzens2011} or
near field nanoantennae \cite{Xiao2012,Dantham2013,Santiago2012}. Sensitivity
enhancements specifically derive from the increased polarisability of
NPs and generation of plasmonic hotspots, which are maximum when
operating at wavelengths close to the LSP resonance of the NP. Near
resonance, however, scattering and absorption losses are also
increased such that resonance quality is degraded hence countering
potential sensitivity gains. A balance must hence be struck. In this
article we therefore consider the hitherto overlooked phenomena of level
repulsion and level crossings in NP coupled high $Q$ microresonators,
arising from reactive and resistive coupling contributions, which in
turn allows optimal NP configurations to be identified. Whilst we consider the case of core-shell NPs coupled to spherical whispering
gallery mode resonators (WGMRs) for simplicity, the underlying physical principles are common
to all resonator geometries and are hence applicable in a broader
sense. 

The structure of this article is as follows. In Section~\ref{sec:hybridisation} we derive and discuss the hybridisation of
photonic and LSP resonant modes, whereby we demonstrate both level
repulsion and crossing. We proceed in
Section~\ref{sec:biosensing} to consider the consequences of this mode
hybridisation within the context of WGM biosensing, in turn allowing the
optimal NP geometry to be found, as is dictated by the detuning of the WGM
and LSP resonances. Concluding remarks are made in Section~\ref{sec:conclusions}.

\section{Mode hybridisation in coupled photonic-plasmonic systems\label{sec:hybridisation}}

Resonances in isolated microcavities have been well studied in the
literature \cite{Bohren1998, Chang1996,Ching1998} and in all cases can be cast into a
secular equation: $|\mathbb{G}| = 0$, where $\mathbb{G}$ is an
appropriate system matrix. Energy can, in general, escape from the
resonator (e.g. via radiation and absorption losses), such that
$\mathbb{G}$ is non-Hermitian and the associated eigenvalues
(and hence resonance frequencies) are complex. For example, for
spherical microresonators $\mathbb{G}$ is a diagonal matrix with
non-zero elements given by $1/\eta_l^{\nu}$ \cite{Foreman2013}, where 
$\eta_l^{\nu}$ are the well known Mie scattering coefficients for
transverse electric (TE, $\nu = M$) or magnetic (TM, $\nu =E$) Mie modes
with polar and azimuthal mode indices $(l,m)$
\cite{Bohren1998}. Accordingly the secular equation decouples in $l$ and
$m$ to yield the more familiar transcendental equation (which is
independent of $m$):
\begin{align}
 \frac{\big[n_I z \, h_l(n_I z) \big]^\prime }{
   h_l(n_I z)} = N \frac{\big[n_{II} z \, j_l(n_{II} z)\big]^\prime}{
   j_l(n_{II} z)}, \label{eq:MieResTrans}
\end{align}
where $N = 1$ or $(n_I/n_{II})^2$ for TE or TM modes respectively, $j_l(x)$
and $h_l(x)$ are the spherical Bessel and Hankel functions of
the first kind, $z=k a$, $k$ is the (complex) vacuum  wavenumber, $a$ is the
resonator radius, prime
denotes differentiation with respect to the argument of the respective
Hankel or Bessel function and $n_I$ ($n_{II}$)
is the refractive index of the surrounding medium (resonator). It
should be noted that Eq.~\eqref{eq:MieResTrans} is conventionally
found by directly considering when the denominator of $\eta_l^\nu$ is
zero. Whilst many solutions to Eq.~\eqref{eq:MieResTrans} can be found,
here we will primarily be interested in bound surface modes,
i.e. WGMs, for which $\mbox{Re}[z] \sim l/n_{II}$ \cite{Lam1992}.

Plasmonic resonances in metallic NPs have similarly seen extensive
research in the literature. Small ($\ll \lambda$) homogeneous NPs, for instance, are often
described using a (complex) polarisability given by the
Clausius-Mossotti relation, such that LSP resonance occurs when the
Fr\"ohlich condition is satisfied \cite{Giannini2011} (as again follows by
considering when the denominator of the polarisability is zero). This treatment is, however,
only approximate and breaks down for larger, or inhomogenous, NPs due to
retardation effects and field non-uniformity. More
generally, the exact polarisability, $\alpha_{\txtpow{Mie}}$, of a
(possibly inhomogenous) spherical NP can be found using generalised
Mie theory, or the $T$-matrix method
\cite{Mackowski1990,Waterman1969} viz.
\begin{align}
\alpha_{\txtpow{Mie}} = -4\pi \frac{ 3i}{2 k^3 \epsilon_1^{3/2}} T_{1}^E
\end{align}
where $T_{1}^E$ is the electric dipole element of the $T$-matrix and
$\epsilon_1 = n_1^2$ is the electric permittivity of the medium surrounding
the NP \cite{Moroz2009}. For homogeneous NPs $T_{1}^E = \eta_1^E$,
whilst for core-shell NPs $T_{1}^E = \kappa_1^E$ (as defined in
Appendix C of \cite{Foreman2013}). Evidently, the methodology adopted
to determine LSP resonances in metallic NPs is formally, and
physically, equivalent to that used for determining resonances in
large microcavities. When considering more general resonator and NP
geometries, similar arguments can also be made, albeit the form of
$\mathbb{G}$ becomes more complicated and separability in $l$ and $m$
(or analogous mode indices)
is likely lost.

When a NP is coupled to a microcavity
the mode structure of the combined system is modified, yielding hybrid photonic-plasmonic modes with
resonance frequencies shifted from the bare resonator
case. Hybrid resonances, can again be found by solution of the
equation $|\mathbb{G}| = 0$, albeit coupling terms must be incorporated
into $\mathbb{G}$. In earlier work, we have derived
the appropriate system matrix describing coupling of arbitrary NPs to
spherical WGMRs \cite{Foreman2013}, whereupon resonance conditions
were found for hybridisation of the electric dipole LSP resonance in a
NP with the TE and TM WGMs. Restricting attention, henceforth, to  a
core-shell NP the resonance conditions are (approximately) given by \cite{Foreman2013}:
\begin{align}\label{eq:res_cond1_dipole}
0 &= \frac{1}{\eta_l^{\nu}}\frac{1}{\kappa_1^{E}}-  \,
\widetilde{\mathcal{U}}_{1m}^{lm}\mathcal{U}^{1m}_{lm} ,
\end{align}
where $\mathcal{U}^{1m}_{lm} = \mathcal{A}^{1m}_{lm}$ or 
$\mathcal{B}^{1m}_{lm}$ are the
translation coefficients for TM and TE modes
respectively, which arise when
relating fields from displaced scatterers. In the absence of coupling,
Eq.~\eqref{eq:res_cond1_dipole} reduces to the 
resonance conditions for the isolated WGMR ($1/\eta_l^{\nu}(k) =0$)
and NP ($1/\kappa_l^{E}(k) =0$) as discussed above, whilst the coupling (embodied in
$\widetilde{\mathcal{U}}_{1m}^{lm}\mathcal{U}^{1m}_{lm}$) 
represents a perturbation to these conditions. It must be emphasised that
symmetry dictates that the quantisation (polar) axis is
defined along the line joining the WGMR and NP centers, hence avoiding
coupling between azimuthal modes. 

Realistically, the broad LSP resonance spectrally overlaps
with multiple polar modes (of both high and low $Q$) within the WGMR, a
point neglected in the derivation of
Eq.~\eqref{eq:res_cond1_dipole}. Upon inclusion of this effect, the
resonance conditions can be written in the form:
\begin{align}\label{eq:res_cond2_dipole}
0 &= 1-\kappa_1^{\nu} \sum_{l=0}^\infty \eta_l^{E} \,
\widetilde{\mathcal{U}}_{1m}^{lm}\mathcal{U}^{1m}_{lm} .
\end{align}
The real and imaginary parts of the complex roots of
Eqs.~\eqref{eq:res_cond1_dipole} and \eqref{eq:res_cond2_dipole} define the resonance frequency and
line-width of the hybrid modes (here termed ``quasi-TM'' and
``quasi-TE'') respectively. Noting that
$\mathcal{A}^{l'm}_{lm} = 0$ for $|m| > 1$ and
$\mathcal{B}^{l'm}_{lm} = 0$ for $|m| \neq 1$, it is evident that
only the low order azimuthal modes hybridise, whilst higher order
azimuthal modes remain unperturbed. Physically this is a consequence of
higher order modes having zero intensity at the site of the NP. 

Greater insight into Eq.~\eqref{eq:res_cond2_dipole} can
be gained by considering the case when the LSP peak lies
spectrally close to the WGM resonance of polar index $L$, as can be
practically realised using the tunability of core-shell NPs
\cite{Halas2002}. Near resonance the scattering coefficients can be represented by complex Lorentzians of
the form
\begin{align}\label{eq:complex_Lorentzian}
\chi_l^{\nu,j}(k) \approx \chi_l^{\nu,j}\left(k_{0,j}\right)
\frac{i \Gamma_{0,j} /2}{ \left(k - k_{0,j}\right) + i \Gamma_{0,j} /2},
\end{align}
where $\chi_l^{\nu,j} = \eta_l^\nu$ or $\kappa_l^\nu$ for $j= 1$ and
$2$ respectively and $k_j = k_{0,j}  - i \, \Gamma_{0,j}/2$ ($k_{0,j},\Gamma_{0,j}
\in \mathbb{R}$) is the complex resonance frequency of an isolated
WGMR ($j=1$) and NP ($j=2$). Note that both the resonance frequency
$k_{0,j}$ and linewidth $\Gamma_{0,j}$ depend on
the mode indices $(\nu,l,m)$, however this dependence has been
suppressed for clarity. We also note for later convenience that $\chi_l^{\nu,j}\!\left(k_{0,j}^\nu\right) \in \mathbb{R}$.

\begin{figure}[t!]
\begin{center}
\includegraphics[width=\columnwidth]{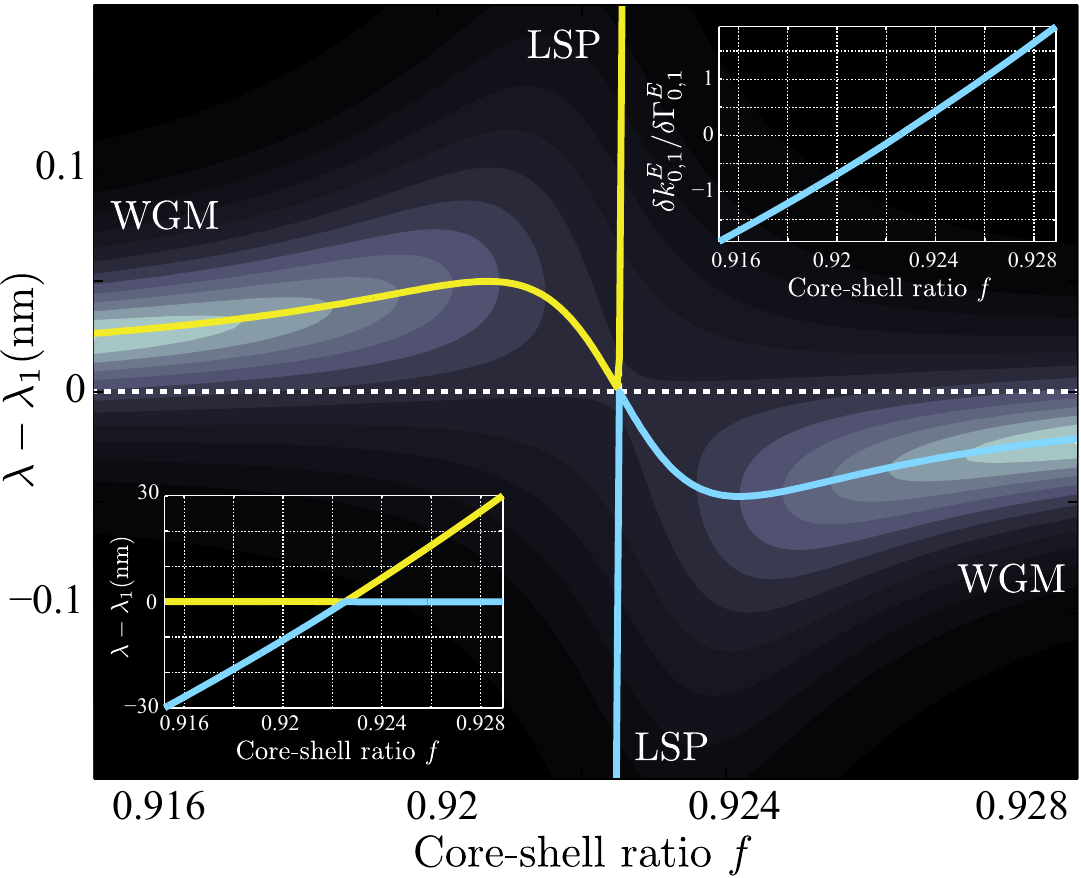}
\caption{(Color online) Normalised extinction spectrum
  vs. ratio of core-shell radii for a coupled WGMR-NP system
  illuminated to excite the $(40,1)$ WGM in a bare
  WGMR. Solid lines show resonance branches defined by
  Eq.~\eqref{eq:transcendental}. Left inset: larger scale view of
  resonance branches. Right inset: ratio of resonance shift, $\delta k^E_{0,1}$, to line
  broadening, $\delta \Gamma^E_{0,1}$, quantifying
  reactive vs resistive coupling. \label{fig:crossing}}
\end{center}
\end{figure}

\begin{figure}[hb!]
\begin{center}
\includegraphics[width=\columnwidth]{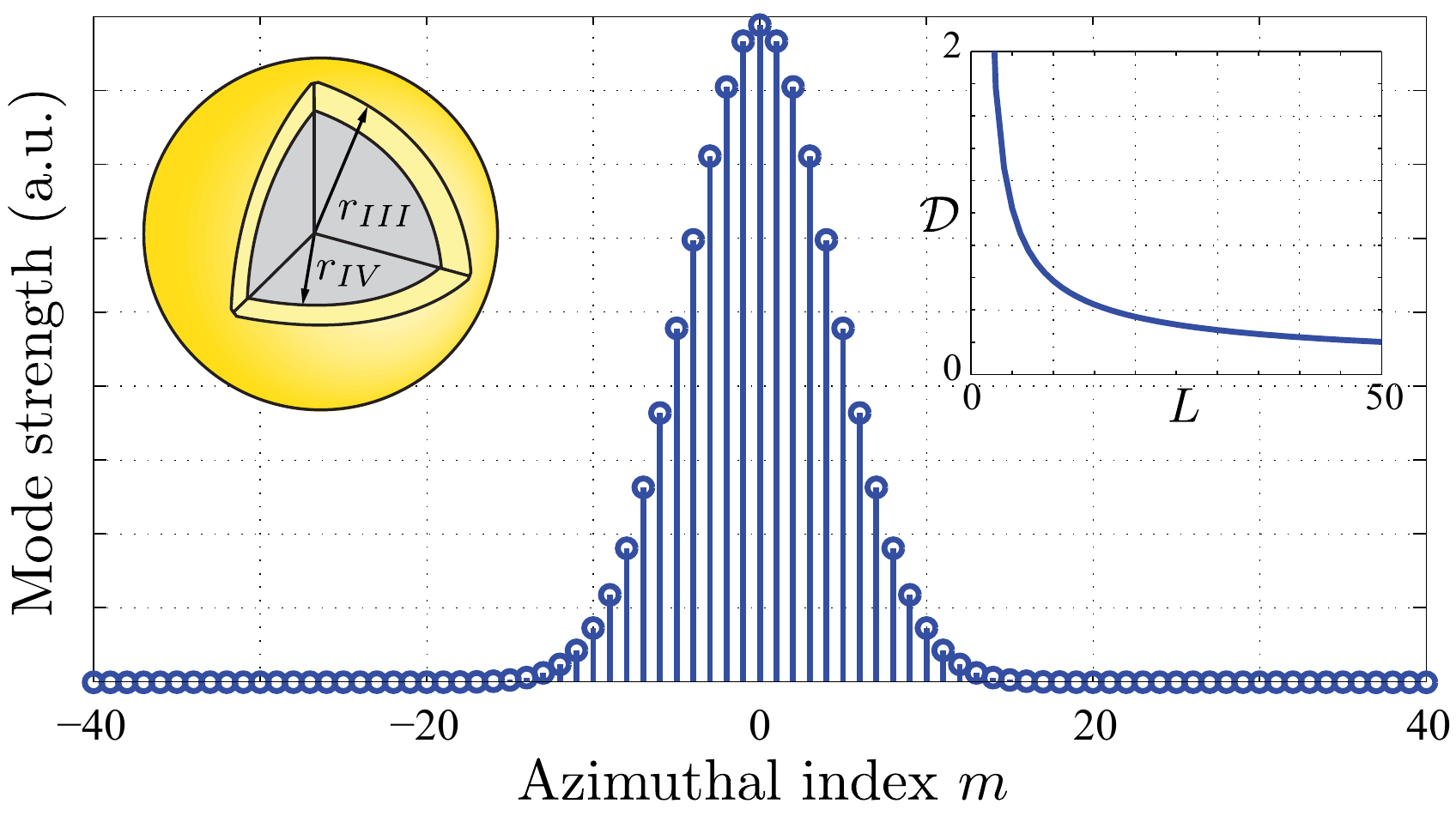}
\caption{(Color online) Mode strengths of degenerate azimuthal modes in a
  fundamental $L=40$ WGM. Left inset: geometry of a core-shell
  NP. Right inset: variation of the ratio $\mathcal{D}=\sum_{|m|\leq1}    \left|D_{m}\right|^2/ \sum_{2\leq|m|\leq L }
\left|D_{m}\right|^2$ with polar index $L$.\label{fig:illum_coeffs}}
\end{center}
\end{figure}

Upon substituting Eq.~\eqref{eq:complex_Lorentzian} into
Eq.~\eqref{eq:res_cond2_dipole} and subsequent rearrangement we arrive
at the quadratic equation :
\begin{align}\label{eq:quadratic}
k_{12}^2 - k_{12}(k_1 + k_2 -
i J) +
k_1 k_2 = -i k_1 J-K
\end{align}
where $k_{12} =k_{0,12}  - i \, \Gamma_{0,12}/2 $ is the complex resonance frequency of the
hybrid mode, 
\begin{align}
J &= \frac{1}{2}\Gamma_{0,2} \,
\kappa_1^{E}   (k_{0,2}) \sum_{l\neq L} \eta_l^{\nu}(k_{0,1}) 
\widetilde{\mathcal{U}}_{1m}^{lm}\mathcal{U}^{1m}_{lm} 
\end{align}
describes the strength of coupling between the LSP and off-resonance
($l\neq L$) modes and
\begin{align}K &=
\frac{1}{4}\Gamma_{0,1} \Gamma_{0,2} \, \eta_L^{\nu}(k_{0,1}) \kappa_1^{E}
  (k_{0,2})
\widetilde{\mathcal{U}}_{1m}^{Lm}\mathcal{U}^{1m}_{Lm} 
\end{align}
describes coupling between the LSP and the on-resonance ($l=L$) WGM. Completing
the square in Eq.~\eqref{eq:quadratic} yields
\begin{align} \label{eq:transcendental}
\delta k_{j} &= \frac{1}{2}\bigg[\left(k_{1} + k_{2}  -
  2k_{j} - i
  J\right)\\ &
\quad\pm \sqrt{ \left(k_{1}
      +k_{2} - i J\right)^2 -
  4 (k_1 k_2 + i k_1 J + K)  } \bigg] \nonumber ,
\end{align}
where $\delta k_{j} = k_{12} - k_{j}$. Eq.~\eqref{eq:transcendental} describes hybridisation
of the LSP and WGM resonances giving rise
to a level crossing. Figure~\ref{fig:crossing} shows the
extinguished power (here letting $J=0$ for illustrative purposes) of a
WGMR-NP system, illuminated by a field which would excite a TM
$(40,1)$ WGM in an isolated WGMR, as a function of the WGM-LSP resonance detuning
(as can be parameterised by the
ratio $f=r_{III}/r_{IV}$ of the core-shell NP radii - see inset of 
Figure~\ref{fig:illum_coeffs}), from which the crossing behaviour
is evident.  We note that there is an anticrossing in the imaginary
part of $k_{j}$, i.e. linewidth. For calculation purposes the WGMR was
assumed to have a refractive index of $n_{II} = 1.59$, radius of
$r_{II}=4$~$\mu$m and to be in air ($n_I
= 1$), such that the $(40,1)$ TM WGM resonance is at a wavelength of 772.459~nm
(the second order radial WGM was considered as was discussed in
\cite{Foreman2013}). Furthermore, a core-shell NP, of outer radius $r_{III} =
32$~nm, with a fused silica ($n_{IV} = 1.48$) 
core and silver shell was taken and assumed
to be bound at the surface of the WGMR. The
permittivity of silver was modelled using a Drude-Lorentz model
whereby
$\epsilon_{III}(\omega) = \epsilon_{\infty} - {\omega_p^2}/{(\omega^2 + i
  \omega\gamma)}$.
Values of $\epsilon_\infty = 3.7$, $\omega_p = 8.9$~eV and
$\gamma = 0.021$~eV were taken. In reality we note that the dielectric function of
metallic layers becomes size dependent when the layer thickness is small in relation to the mean-free path
of electrons \cite{Kreibig1974}. In turn, this causes a broadening of the LSP resonance
due to additional surface collisions of the electrons in the metal,
however, use of a more realistic permittivity function, such as that
presented in \cite{Miao2010}, does not affect
the physical conclusions of this article, such that a Drude-Lorentz
model was adopted for simplicity. 

Solid lines in
Figure~\ref{fig:crossing} depict the resonance branches defined by
Eq.~\eqref{eq:transcendental}, which swap nature as the resonance
detuning changes sign. Due to the small size of the NP, resonance shifts
of the WGM-like resonance are of the order of picometers, such
that the LSP-like branches appear as near vertical
lines in Figure~\ref{fig:crossing} (see left inset for a larger scale view). Strong extinction is
not seen on the LSP branches since we assume the LSP is only
excited through leakage from the WGMR. From Figure~\ref{fig:crossing}
it can be seen that for larger detunings the WGM and LSP modes are
mutually repelled, however at small detunings the repulsion effect
diminishes, the WGM resonance broadens due to increased scattering and absorption losses arising
from stronger (resistive) coupling to the NP, and ultimately the
resonance branches cross. An optimal NP core-shell ratio can hence be
identified whereby reactive resonance shifts are maximised, as shall
be determined analytically in what follows.

\section{Optimal resonance shifts in WGM biosensing \label{sec:biosensing}}

WGM based sensors operate by what has become known as the reactive
 sensing principle \cite{Arnold2003}. Specifically, the spectral
position of a WGM resonance is shifted when a
molecule enters the near field of the WGMR, by an amount proportional to the
polarisability of the particle. Typical shifts of WGM resonances in
biosensing experiments are, however, on the
femtometer scale, such that the shift of the $m=1$ WGM mode
depicted in Figure~\ref{fig:crossing} is comparatively
large. However, due to the choice of polar axis, excitation of a
fundamental WGM, commonly used for sensing and considered as a single mode,
must be represented by a superposition of degenerate azimuthal
modes \cite{Deych2009}. Since only the low order azimuthal modes
are perturbed (hence lifting the degeneracy) the total
extinction spectrum is formed by superposition of $2 L - s$
unshifted resonances ($s=2$ or $1$ for quasi-TM and -TE modes respectively)
with $s+1$ shifted modes viz.
\begin{align}\label{eq:extinct}
C_{\txtpow{ext}}(k) = C_0\sum_{l=0}^\infty\sum_{m = - l}^l A_{m} \frac{{\Gamma_{0,12}^2}}{4(k-k_{0,12})^2 + {\Gamma_{0,12}^2}}.
\end{align}
Here a Lorentzian mode profile has again been adopted and the
dependence of each quantity on $(\nu,l,m)$ is again suppressed. $C_0$ is an
appropriate normalisation constant and $A_{m}$ denotes
the strength of each azimuthal mode and is dependent on the
illuminating field \cite{Foreman2013}. Summation over $l$ in
Eq.~\eqref{eq:extinct} accounts for additional
non-resonant contributions to the total extinction spectrum from
$l\neq L$ polar modes. In practise the summation limits can be
truncated since realistic coupling schemes for WGMRs excite a
limited range of polar modes \cite{Serpenguzel1997}. 

To determine the apparent shift of the measured WGM resonance
lineshape (as
dictated by the superposition of both shifted and unshifted azimuthal
modes) the spectral position, $k_{\txtpow{max}}$ of the maximum in Eq. ~\eqref{eq:extinct} must be found. Therefore, we
differentiate Eq.~\eqref{eq:extinct} with respect to $k$ and equate
the result to zero as per standard theory, ultimately yielding
\begin{align}
0 =&\sum_{|m| \leq 1} A_{m}  \frac{{\Gamma  _{0,12}^2} (k_{\txtpow{max}}-k_{0,1}  - \delta k_{0,1} ) }
{\left[4(k_{\txtpow{max}}-k_{0,1}  - \delta k_{0,1} )^2 +
    {\Gamma_{0,12}^2 } \right]^2} \nonumber\\
&+ \sum_{2\leq|m|\leq L} A_{m}  \frac{{\Gamma_{0,1}^2 } (k_{\txtpow{max}}-k _{0,1}) }
{\left[4(k_{\txtpow{max}}- k _{0,1})^2 + {\Gamma_{0,1}^2} \right]^2} \label{eq:spec_diff}
\end{align}
where $\delta k_{0,j}  = \mbox{Re}[\delta k_{j}  ]$ and the non-resonant contributions are assumed to vary negligibly with wavenumber near
the WGM resonance.
Broadening of a single $|m|\leq 1$ mode is well described by Larmor's
formula for an oscillating dipole \cite{Jackson1998}, however, for our
purposes it is safe to neglect broadening of the
$|m|\leq1$ modes in Eq.~\eqref{eq:spec_diff} (only), i.e. let $\Gamma_{0,12} \approx
\Gamma_{0,1}$ (as has also been confirmed via full Mie scattering calculations). Further dropping the
$\delta k_{0,1} $ term in the denominator, yields
the peak position relative to the isolated fundamental WGM resonance as:
\begin{align}\label{eq:total_max}
k_{\txtpow{max}}-k_{0,1} &\lesssim \frac{\sum_{|m|\leq1} \left|D_{m}\right|^2 \,  \delta
  k_{0,1} }{\sum_{2\leq|m|\leq L }  \left|D_{m}\right|^2 } \leq
\sum_{|m|\leq1}  \frac{\delta
  k_{0,1}  }{2 L +1} ,
\end{align}
where $D_m
= \sqrt{(2 L)! /[(L+m)!(L-m)!]}$ derive from the Wigner $D$ functions \cite{Deych2009}. The first
inequality in \eqref{eq:total_max} is satisfied when the
NP lies in the plane of the fundamental mode, and follows because
coupling is strongest in this
case. The latter (weaker) inequality follows by noting the $m=0$ mode is strongest for a
fundamental mode, with mode strength decreasing with increasing $|m|$
(see Figure~\ref{fig:illum_coeffs}).
\begin{figure}[t!]
\begin{center}
\includegraphics[width=\columnwidth]{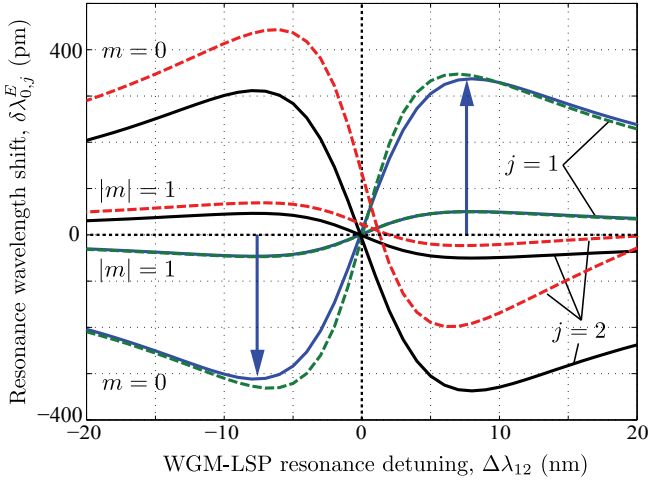}
\caption{(Color online) Resonance shifts of $|m| \leq 1$ quasi-TM modes
  for $j=1$  (blue and green) and $j=2$ (black and red). Solid
  (dashed) curves
  denote shifts calculated using Eq.~\eqref{eq:hybrid_shift2}
  (Eq.~\eqref{eq:res_cond1_dipole}). Arrows denote detunings of $\pm
  \Gamma_{0,2}/2$ (see Eq.~\eqref{eq:approx_opt2}). \label{fig:mode_shift}}
\end{center}
\end{figure}  

Eq.~\eqref{eq:total_max} illustrates that to maximise the
total apparent shift of the WGM resonance in a typical WGM biosensing experiment, we must maximise the sum of the
shifts of the individual $|m|\leq1$ modes. 
We perform this optimisation by varying the
ratio of the core-shell radii, $f$, at fixed $k_{0,1}$. Noting that WGMR-NP coupling
is weak we first approximate
Eq.~\eqref{eq:transcendental} further, yielding
\begin{align}\label{eq:hybrid_shift2}
\delta k_{0,j}  = (-1)^j\mbox{Re}\left[ \frac{K  + 2 i k_1 
    J }{k_{1}  -k_{2} + i J } \right].
\end{align}
Eq.~\eqref{eq:hybrid_shift2} is a transcendental equation in
$k_{0,12}$, however, as an approximation the translation
coefficients (implicit in $J$ and $K$) are evaluated at the isolated WGM resonance frequency
$k_{0,1}$. Resonance shifts then follow by direct
evaluation of Eq.~\eqref{eq:hybrid_shift2}. An example of such a
calculation is shown in Figure~\ref{fig:mode_shift} for the $j=1$
(solid lines) and $|m|\leq
1$ quasi-TM modes, as a function of LSP-WGM detuning $\Delta k_{12} =
k_{0,1}  - k_{0,2}$, with $J =0$ (i.e. neglecting coupling to
off-resonant modes of the WGMR). Shifts of the $j=2$ (LSP) resonance are
also depicted in Figure~\ref{fig:mode_shift} for reference. Dashed
curves in Figure~\ref{fig:mode_shift} describe shifts
determined by exact numerical solution of
Eq.~\eqref{eq:res_cond1_dipole} and are seen to be shifted
to the left relative to the approximate plots, albeit the shift is negligible for
the WGM-like modes. Slight narrowing of the transition region is
also evident. Larger disparities between the approximate and exact
curves for $j=2$ modes are seen.

Regarding Figure~\ref{fig:mode_shift} a number of details are worthy
of mention. Primarily, as also seen above, clear turning points are exhibited in the $j=1$
plots, indicating that by judicious choice of NP geometry the shift of the WGM resonance can indeed be
maximised. It is further noted that turning points for the $|m|\leq1$
modes lie in close proximity. Consequently,
maximising the sum of shifts is practically equivalent to maximising the individual
shifts, since this is achieved for near identical NP geometries (and
certainly within current fabrication tolerances). The nature of the
apparent WGM shift, however, differs with the sign of the WGM-LSP
detuning. Specifically, when the isolated LSP
resonance lies at shorter (longer) wavelengths than the WGM resonance, the
shift is towards the red (blue) end of the spectrum. 

Determination of the optimal NP geometry requires locating the turning
points of Eq.~\eqref{eq:hybrid_shift2}. Unfortunately, $\kappa_1^E$ depends on $f$, complicating the
process further. Nevertheless, over the wavelength
ranges considered the variation is weak, such that when evaluating $\kappa_1^E$,
$f$ can be set at a sensible value (here taken such that
$\Delta k_{12} = 0$). Performing the
stationary point analysis then gives an estimate
of the optimal detuning as:
\begin{align}
\Delta k_{12}^{\txtpow{opt}} =&
\bigg[\mbox{Im}[\widetilde{K} ]\Delta\Gamma_{12} -
  \mbox{Re}[K  {J }^*] + |J |^2\Gamma_{0,1} \nonumber\\
& \pm \sqrt{|\widetilde{K} |^2(\Delta\Gamma_{12} -
    2\mbox{Im}[J ])^2}\bigg] \Big/ \mbox{Re}[2\widetilde{K} ] \label{eq:approx_opt}
\end{align}
where $\widetilde{K}  = K  + 2i k_1  J $ and  $\Delta
\Gamma_{12} = \Gamma_{0,1} - \Gamma_{0,2}$.

To proceed, we examine the coupling terms,
$K $ and $J $, further. In Appendix B of \cite{Foreman2013} it
was shown that 
\begin{align}
\mathcal{A}^{1m}_{lm}  \sim  \frac{(l+m)!}{l!}h_{l-1}(z) +  \frac{(-1)^m \,l!}{(l-m)!}h_{l+1}(z)
\end{align}
where here $z = n_I k r_{\txtpow{NP}}$ and $r_{\txtpow{NP}}$ denotes the NP
displacement relative to the WGM. Using the recurrence relations
\cite{Abramowitz1970} for
$h_l(z)$:
\begin{align}
(2l+1) h_l^\prime(z) &= l h_{l-1}(z)  - (l+1) h_{l+1}(z)\\
\frac{2l+1}{z}h_l(z)&=h_{l-1}(z) + h_{l+1}(z)
\end{align}
 and noting $k
  r_{\txtpow{NP}} \gg 1$ it follows that:
\begin{align}
\mbox{arg}\left[\widetilde{\mathcal{A}}_{1,\pm 1}^{l,\pm
    1}\mathcal{A}^{1,\pm 1}_{l,\pm1}\right]
&= 2\, \mbox{arctan} \left[ y_l'(z) / j_l'(z)  \right] = 2\varphi_l(z)\\
\mbox{arg}\left[\widetilde{\mathcal{A}}_{1,0}^{l,0}\mathcal{A}^{1,0}_{l,0}\right]
&= 
 2\, \mbox{arctan} \left[ y_l(z) / j_l(z)  \right]  = 2\vartheta_l(z)
\end{align}
where $y_l(z)$ is the spherical Bessel function of the second kind. Similarly,
$\mbox{arg}[\widetilde{\mathcal{B}}_{1m}^{lm}\mathcal{B}^{1m}_{lm}]
= 2\vartheta_l(z)$ for $|m| = 1$. The phase functions $\vartheta_l(z)$ and
$\varphi_l(z)$ are approximately $ \pm \pi/2$ for $z$ smaller than the
first zero of $h_l(z)$ and $h_l'(z)$ respectively. Noting the asymptotic expansions
for the zeros of $h_l^{(\prime)}(z)$ \cite{Abramowitz1970} and the WGM
resonance frequencies \cite{Lam1992} and also observing coupling only
occurs when the NP is on (or near) the WGMR surface (i.e. $r_{\txtpow{NP}}\approx r_{II}$), we find that
$K $ is predominantly real when 
\begin{align}\label{eq:reactive_ineq}
\left({n_{II}}-n_I \right)L >  2^{-1/3}n_I\alpha_i - n_{II}\beta_1^{(\prime)} ,
\end{align}
where $\alpha_i$ denotes the $i$th negative zero of the Airy
function (dictating the radial order of the WGM) and
$\beta_1^{(\prime)}$ denotes the first zero of
$h_l^{(\prime)}(z)$.  Inequality \eqref{eq:reactive_ineq} is
easily satisfied in practice for high $Q$ WGMs. For $|\Delta k_{12}| > |\Delta \Gamma_{12}|/2$, real $K $ implies 
reactive coupling, i.e. the increase in the
half-width of the WGM-like resonance is smaller than the resonance
shift (see right inset of Figure~\ref{fig:crossing}). If, however, $|\Delta k_{12}| < |\Delta
\Gamma_{12}|/2$, i.e. small detunings, coupling becomes resistive even for purely real
$K $, due to the losses of the isolated resonances
(as seen from Eq.~\eqref{eq:hybrid_shift2} and its broadening
counterpart). In contrast, the
$J $ term is dominated by its imaginary part since the
$\eta_{l\neq L}^{E,1}$ terms
are strongly imaginary away from resonance. Eq.~\eqref{eq:approx_opt}
can therefore be approximated as
\begin{align}\label{eq:approx_opt2}
\Delta k_{12}^{\txtpow{opt}} \approx& \frac{|\widetilde{K} | |\Delta\Gamma_{12} -
    2\,\mbox{Im}[J ]|}{2\,\mbox{Re}[\widetilde{K} ]} \approx \left|\frac{\Gamma_{0,2}}{2} +
J \right|,
\end{align} 
where the last
step follows for high $Q$ WGMs whereby $\left|\Delta
  \Gamma_{12}\right| \approx \Gamma_{0,2}$. When coupling to $l\neq L$
polar modes is negligible, Eq.~\eqref{eq:approx_opt2} states that the
optimal WGM resonance shifts are obtained when the LSP of the
NP is detuned by half the line-width of the LSP. Spectral
overlap of the LSP with other modes within the WGMR, however, produces
a shift in this optimal detuning. Induced shifts can be either to
greater or smaller detunings, however, noting that the $l\neq L$ modes
are
excited off-resonance, the effective shift is small. Optimal detunings
predicted using Eq.~\eqref{eq:approx_opt2} are shown in Figure~\ref{fig:mode_shift} by the blue
arrows. Eq.~\eqref{eq:approx_opt2} is thus seen to provide a good
rule for optimising core-shell NP geometries and is the main result of
this article. Slight
numerical differences arise due to the approximations
taken.

\section{Discussion \label{sec:conclusions}}
Implicitly, throughout this work it has been assumed that
shifts of the $j=1$, $|m|\leq 1$ resonances are small relative to their line-widths. If this criterion is not satisfied mode splitting can
occur. From
Eq.~\eqref{eq:total_max} the maximum possible
resonance shift is seen to decrease with increasing $L$ (and hence increasing WGMR
radius), as illustrated in Figure~\ref{fig:illum_coeffs} which plots the ratio
$\mathcal{D}=\sum_{|m|\leq1}    \left|D_{m}\right|^2/ \sum_{2\leq|m|\leq L }
\left|D_{m}\right|^2$. Since the line-width of WGM resonances,
however, decreases with WGMR radius as $\sim[ r_{II}
\,y_L(k r_{II})]^{-2}$ \cite{Lam1992}, mode splitting
is easier to observe in larger WGMRs, albeit this is limited by
loss mechanisms, such as NP scattering and absorption
\cite{Ozdemir2011}. Splitting is hence only observable for a
restricted range of NP sizes and is not predicted in this work.

Given that the treatment detailed in this article is largely
mathematical, it is important to give a more quantitative estimate of
the enhancement of the resonance shifts that may be expected in a
typical experiment. Assuming a silica-silver core-shell NP of 55~nm
outer radius, we estimate that at the optimal detuning an approximate
90-fold enhancement of the WGM resonance shift can be obtained, as
compared to the WGM shift resulting from perturbation by a 
dielectric sphere of the same dimensions and with a refractive index of 1.5. Use of
a gold shell NP reduces the obtainable enhancement to $\sim \times
22$ due to a lower quality LSP resonance. Noting that the maximum near field
enhancement (as defined in \cite{Tanabe2008}) arising from a
plasmonic NP scales as the
square of the polarisability \cite{Foreman2012}, we estimate an
enhancement factor of $\sim 150$ for both gold and silver. A more considered
choice of NP materials allows even greater near field enhancements to
be achieved \cite{Tanabe2008}. Surface roughness of the NP has
also recently been shown to have a significant influence on achievable
near field
enhancements \cite{Dantham2013}. Design of the NP to achieve coupling between the WGM and
higher order LSP resonances, such as the electric quadrupole resonance,
would furthermore allow larger enhancements due to the higher $Q$ of the LSP
resonances. Use of more complicated NP geometries allowing greater
flexibility and control of the NP spectral properties
\cite{Giannini2011} thus presents an important avenue for
plasmon enhanced WGM biosensing, however, binding orientation of
asymmetric NPs then also affects the resonance shifts and near field
enhancements that can be observed \cite{Foreman2013}. 

To summarise, in this article we have considered the use of core-shell
NPs as near field nanoantennae or analyte labels for sensing
purposes. Improvement of WGM sensors was shown to be possible by
tuning the LSP resonance to lie approximately half the LSP (on
resonance) line-width from the WGM resonance, whereby an optimal
balance of line broadening and shifts is achieved. Sensitivity
enhancements were shown to derive from mode hybridisation such that level repulsion
played a key role. Criteria under which WGMR-NP coupling is chiefly reactive
were also given.

The authors would like to acknowledge financial support from the Max Planck Society.

\end{document}